\documentclass[pra,twocolumn,superscriptaddress,10pt,noshowpacs]{revtex4}
\usepackage[english]{babel}
\usepackage[T1]{fontenc}
\usepackage[utf8]{inputenc}
\usepackage{graphicx,epstopdf}
\usepackage{amsmath}

\usepackage{amsfonts}
\usepackage{bbm}
\usepackage{amssymb}
\usepackage{color}
\usepackage{latexsym}
\usepackage{caption}
\usepackage{subcaption}
\usepackage{times,txfonts}

\begin{document}

\title{Charged black string bounce and its field source}

\author{A. Lima}
\affiliation{Universidade Federal do Cear\'{a}, Fortaleza, Cear\'{a}, Brazil}
\email{arthur.lima@fisica.ufc.br}

\author{G. Alencar}
\email{geova@fisica.ufc.br}
\affiliation{Universidade Federal do Cear\'{a}, Fortaleza, Cear\'{a}, Brazil}

\author{R. N. Costa Filho}
\email{rai@fisica.ufc.br}
\affiliation{Universidade Federal do Cear\'{a}, Fortaleza, Cear\'{a}, Brazil}

\author{R. R. Landim}
\email{renan@fisica.ufc.br}
\affiliation{Universidade Federal do Cear\'a (UFC), Departamento de F\'isica,\\ Campus do Pici, Fortaleza - CE, C.P. 6030, 60455-760 - Brazil.}

\date{\today}

\begin{abstract}
This work builds upon the previous article \cite{Lima:2022pvc} and explores the solution of the charged black string introduced in \cite{Lemos:1995cm}. The black bounce regularization method, based on the Simpson-Visser solution, is employed by transforming the radial variable using $r\rightarrow \sqrt{r^2+a^2}$. The regular charged black string metric is defined, and the properties of event horizons, surface gravity, and Hawking temperature are investigated. The behavior of curvature quantities, including curvature invariants and tensors, is examined to verify the absence of singularities when $a\neq 0$. The Einstein equation for the energy-momentum tensor is solved, and the null energy condition is analyzed for the obtained solution. The sources of this solution are evaluated, combining a scalar field with nonlinear electrodynamics. However, unlike other works, an electric field is considered instead of a magnetic field. Finally, the study calculates the possibility of stable or unstable circular orbits for massive and massless particles.      
\end{abstract}

\maketitle

\section{Introduction}\label{Intro}

In recent years, Black holes have gained significant attention due to technological advancements, which are evident in the detection of gravitational waves by projects such as LIGO and VIRGO, as well as the first images of supermassive black holes \cite{LIGOScientific:2016aoc, EventHorizonTelescope:2022wkp, EventHorizonTelescope:2019dse, LIGOScientific:2017vwq}. A possible, less common solution is to consider cylindrical symmetry. This type of solution is particularly important for cosmology because in the study of the evolution of the Universe, in the phase transitions that may have occurred after the \textit{big bang}, the so-called topologically stable defects are studied, such as cosmic strings \cite{Vilenkin:1984ib}. The cylindrical counterparts of black holes are known as black strings \cite{Lemos:1995cm}.

Black holes and black strings exhibit singularities in their curvature. These singularities cannot be ignored when dealing with evaporation through Hawking radiation, particularly in the final stages of evaporation. Providing an accurate description of spacetime under these conditions is crucial in addressing certain questions that involve the relationship between General Relativity and Quantum Mechanics \cite{Carballo-Rubio:2018pmi}. However, it is possible to develop regular solutions, which can circumvent these issues and shed light on evaporation problems \cite{Hayward:2006, bardeen:1968non, Bambi:2013ufa, Bronnikov:2005gm, Bronnikov:2000vy}. 

In particular, Simpson and Visser \cite{Simpson:2018tsi}  proposed a regular solution by changing $r^2\rightarrow r^2+a^2$ in the Schwarzschild solution. Soon later, the same regularization procedure was applied for Reisnerr-Nordstrom, Kerr-Newman black holes  and many solutions and properties were studied \cite{Simpson:2021vxo,Bambhaniya:2021ugr,Terno:2022qot,Junior:2022fgu,Islam:2021ful, Nascimento:2020ime, Tsukamoto:2020bjm, Ovgun:2020yuv, Chataignier:2022yic, Bronnikov:2022bud, Stuchlik:2021tcn, Churilova:2019cyt, Yang:2022ryf, Vagnozzi:2022moj, Lobo:2020ffi}. An important question is what source could generate this kind of solution, which requires exotic sources. Using nonlinear electrodynamics (NED) to represent regular black hole solutions is quite common \cite{Rodrigues:2023vtm, Bronnikov:2021uta}. In another direction, using a phantom field to represent traversable wormholes is also common. However, for "black bounce" solutions, using these sources individually is impossible, and  the two types of solutions must be considered \cite{Bronnikov:2021uta, Bakhtiarizadeh:2023mhk}.

The Simpson-Visser regularization was recently applied to the black string solution \cite{Lima:2022pvc}. In this case, the regularization $r^2\rightarrow r^2+a^2$ was applied to the metric defined in Ref. \cite{Lemos:1994xp}. However, up to now, the regularization of the charged case \cite{Lemos:1995cm}, analogous to Reissner-Nordström, has not been considered.  Therefore, we aim to apply the Simpson-Visser method to the charged case, conducting similar analyses to what was done in \cite{Franzin:2021vnj}. 

\section{Black-bounce charged black string}\label{Sec-2}
Let's begin by defining the solution for a charged static black string, which can be found in \cite{Lemos:1995cm}. In this solution, we consider a spacetime with cylindrical symmetry, independent of time, and in the presence of an electrostatic potential $h(r)$. The metric is given by:
\begin{equation}\label{3}
    ds^2=-f(r)dt^2+\frac{dr^2}{f(r)}+r^2d\varphi^2+\alpha^2r^2dz^2,
\end{equation}
where 
\begin{equation}\label{4}
    f(r)=\left(\alpha^2r^2-\frac{b}{\alpha r}+\frac{c^2}{\alpha^2r^2}\right).
\end{equation}
Here, $\alpha^2=-\Lambda/3$, $b=4\mu$, where $\mu$ represents the mass per unit length of the black string, and $c^2=4\lambda^2$, where $\lambda$ represents the linear charge density of the black string. The electrostatic potential is given by $h(r)=-2\lambda/(\alpha r)+h_0$, where $h_0$ is an arbitrary constant \cite{Lemos:1995cm}.

However, this solution is not regular as it has a singularity at the origin, as indicated by the Kretschmann scalar: 
\begin{equation}\label{9}
    R^{\mu\nu\lambda\rho}R_{\mu\nu\lambda\rho} = 24\alpha^4\left(1+\frac{b^2}{2\alpha^6r^6}\right)-\frac{48c^2}{\alpha^3r^7}\left(b-\frac{7c^2}{6\alpha r}\right).
\end{equation}

Our current objective is to regularize this solution using the Simpson-Visser method, as done in \cite{Simpson:2018tsi, Franzin:2021vnj}. To achieve this, we need to modify the metric defined in (\ref{3}) to obtain the "black-bounce" solution:
\begin{equation}\label{10}
ds^2=-g(r)dt^2+\frac{dr^2}{g(r)}+(r^2+a^2)d\varphi^2+\alpha^2(r^2+a^2)dz^2,
\end{equation}
where
\begin{equation}\label{11}
g(r)=\left(\alpha^2(r^2+a^2)-\frac{b}{\alpha \sqrt{r^2+a^2}}+\frac{c^2}{\alpha^2(r^2+a^2)}\right).
\end{equation}

It is straightforward to verify that this solution reduces to the one defined in (\ref{3}) as $a\rightarrow 0$. Furthermore, in the asymptotic limit, both solutions are not asymptotically flat due to considering a nonzero cosmological constant.

\subsection{\textit{Event Horizons and Surface Gravity}}
\label{EHSG}
The objective of this section is to determine the Hawking temperature. First, we need to determine the positions of the event horizons defined by $g(r)=0$. We obtain 
\begin{equation}\label{a}
r_{\pm}=S_2 \sqrt{b^{2/3}\frac{\left(\sqrt{s} + S_1 \sqrt{2\sqrt{s^2-4q^2}-s}\right)^2}{4\alpha^2}-a^2},
\end{equation}
where $S_1=\pm1$ and $S_2=\pm 1$. The signs present in $S_1$ refer to the outer and inner horizons, similar to the non-regular solution ($a=0$). The signs present in $S_2$ are related to the "universe" we are considering, where the plus sign refers to the upper universe and the minus sign refers to the lower universe, as in traversable wormholes \cite{Morris:1988cz}.

Assuming that there are two horizons for the solution with $a=0$, when analyzing $r_{\pm}$, we must consider several possibilities: $a>\overline{r}_{+}$, $\overline{r}_{-}<a<\overline{r}_{+}$, $a<\overline{r}_{-}$, where $\overline{r}_{\pm}$ refer to the horizons of the non-regular solution ($a=0$). For $a>\overline{r}_{+}$, it can be easily seen from equation (\ref{a}) that the values of $r_{\pm}$ will be complex, both for the inner and outer horizon. Therefore, the horizons disappear and we have a solution of a traversable wormhole. In the case where $a<\overline{r}_{-}$, both horizons are real, indicating a black hole solution with two horizons. In the case where $\overline{r}_{-}<a<\overline{r}_{+}$, only the outer horizon exists, as the solution for the inner horizon would be complex. Therefore, we have a black hole with only one horizon.

Now, let's determine the Hawking Temperature using $T_{H}=-\partial_1g_{00}(r_+)/(4\pi)$ \cite{Alencar:2018vvb}. We get
\begin{equation}
T_{H}=T_{HCS}\sqrt{\frac{r_+^2}{r_+^2+a^2}}.
\end{equation}
with $T_{HCS}=\kappa_{HCS}/2\pi$, where $\kappa_{HCS}$ is the surface gravity of the usual charged solution in which $\kappa_{HCS}=\alpha^2\overline{r}_+ + \frac{b}{2\alpha \overline{r}_+^2}-\frac{2c^2}{\alpha^2\overline{r}_+^3}$.

Now, let's evaluate the curvature generated by this solution to assess its regularity, as well as determine the basic tensors that will be used to solve the Einstein field equations. This analysis will provide insights into the energy-momentum tensor and the energy conditions of the system.

\subsection{\textit{Curvature Invariants}}
\label{CI}
Our analysis in this section aims to verify the absence of singularities in the regularized solution of the charged black string. Since this solution is static, it is sufficient for our analysis to study the Kretschmann scalar, as its finiteness at any point in space guarantees that all the orthonormal components of the Riemann curvature tensor will also be finite, ensuring the regularity of our solution \cite{Simpson:2023apa}.

The Kretschmann scalar of the solution defined in (\ref{10}) can be written as:
\begin{eqnarray}
    \nonumber &&R^{\mu\nu\lambda\rho}R_{\mu\nu\lambda\rho} = \frac{1}{\alpha^4(r^2+a^2)^6}[4(r^2+a^2)^2(14c^4- \\
    \nonumber &&12\alpha bc^2\sqrt{r^2+a^2}+3b^2\alpha^2(r^2+a^2)+6\alpha^8(r^2+a^2)^4)+ \\
    \nonumber &&a^4(108c^4+33b^2\alpha^2(r^2+a^2)+12\alpha^8(r^2+a^2)^4-\\ \nonumber &&8c^2(14b\alpha\sqrt{r^2+a^2}+\alpha^4(r^2+a^2)^2))-4a^2(r^2+a^2)(38c^4+\\ 
    \nonumber &&9b^2\alpha^2(r^2+a^2)+3b\alpha^5(r^2+a^2)^{5/2}+6\alpha^8(r^2+a^2)^4-\\
     &&c^2(35b\alpha\sqrt{r^2+a^2}+4\alpha^4(r^2+a^2)^2))]. \label{12}
\end{eqnarray}

Now, we need to evaluate it at the origin, $r=0$:
\begin{eqnarray}
    \nonumber R^{\mu\nu\lambda\rho}R_{\mu\nu\lambda\rho} &=& \frac{3\alpha^4(3r_{HNS}^6+4a^6-4r_{HNS}^3a^3)}{a^6} +\\
     &&\frac{12c^4}{\alpha^4a^8}+\frac{4c^2(2a^3-5r_{HNS}^3)}{a^7},
\end{eqnarray}
where $r_{HNS}=b^{1/3}/\alpha$ is the position of the events horizon of the usual neutral solution \cite{Lemos:1994xp}. We observe that as $r \rightarrow 0$, the Kretschmann scalar is finite if $a \neq 0$, which guarantees the regularity of our solution. The condition $a \neq 0$ will always hold for our analysis, as otherwise, we would recover the original solution from (\ref{3}). However, our objective is precisely to evaluate new solutions that are not the original one.

We can also analyze other invariants such as the Ricci scalar, for example:
\begin{equation}\label{13}
R = -12 \alpha^2 + \frac{3a^2\alpha^2(r_{HNS}^3+2(r^2+a^2))^{3/2})}{ \overline{r}^{5/2}}-\frac{2a^2c^2}{\alpha^2(r^2+a^2)^3},
\end{equation}

where $r_{HSN}\equiv b^{1/3}/\alpha$ is the position of the event horizon of the non-regular solution of the uncharged black string. Thus, as $r\rightarrow 0$:
\begin{equation}\label{14}
R = -12 \alpha^2 + \frac{3\alpha^2(r_{HNS}^3+2a^3)}{a^3}-\frac{2c^2}{\alpha^2a^4},
\end{equation}
therefore, no singularities at the origin.

\subsection{\textit{Curvature Tensors}}
\label{CT}

Having evaluated some curvature invariants, let us determine the main tensors that involve the curvature of our spacetime, specifically the Riemann and Ricci tensors. Starting with the Riemann tensor, its non-zero and independent components are given by:
\begin{eqnarray}
    \nonumber R^{01}{}{}_{01} &=& \frac{\alpha^2[2r_{HNS}^3(r^2+a^2)-3a^2r_{HNS}^3-2(r^2+a^2)^{5/2}]}{2(r^2+a^2)^{5/2}}+ \\
    &&\frac{(a^2-3r^2)c^2}{\alpha^2(r^2+a^2)^3}; \label{15}\\  
    \nonumber R^{02}{}{}_{02}&=&R^{03}{}{}_{03}= -\frac{\alpha^2r^2[r_{HNS}^3+2(r^2+a^2)^{3/2}]}{2(r^2+a^2)^{5/2}}+\\
    &&\frac{r^2c^2}{\alpha^2(r^2+a^2)^3}; \label{16}\\
    \nonumber R^{12}{}{}_{12}&=&R^{13}{}{}_{13}=\frac{\alpha^2[2r_{HNS}^3a^2-r^2r_{HNS}^3-2(r^2+a^2)^{5/2}]}{2(r^2+a^2)^{5/2}}+\\
    && \frac{(r^2-a^2)c^2}{\alpha^2(r^2+a^2)^3}; \label{17}\\
    \nonumber R^{23}{}{}_{23}&=& -\frac{\alpha^2r^2[(r^2+a^2)^{3/2}-r_{HNS}^3]}{(r^2+a^2)^{5/2}}-\\
    &&\frac{r^2c^2}{\alpha^2(r^2+a^2)^3}. \label{18}
\end{eqnarray}

From these, we can determine the non-zero and independent components of the Ricci tensor:
\begin{eqnarray}
    \nonumber &R^{0}{}_{0} =& \frac{\alpha^2[a^2(4(r^2+a^2))^{3/2}-r_{HNS}^3)-6(r^2+a^2)^{5/2}]}{2 (r^2+a^2)^{5/2}}-\\
    &&\frac{(r^2-a^2)c^2}{\alpha^2(r^2+a^2)^3}; \label{19}\\
    \nonumber &R^{1}{}_{1} =& \frac{\alpha^2[3a^2r_{HNS}^3-6(r^2+a^2)^{5/2}]}{2 (r^2+a^2)^{5/2}}-\\
    &&\frac{c^2}{\alpha^2(r^2+a^2)^2};\label{20}\\
    \nonumber &R^{2}{}_{2} =& R^{3}{}_{3} = \frac{a^2\alpha^2}{(r^2+a^2)}\left(\frac{r_{HNS}^3}{(r^2+a^2)^{3/2}}+2\right)-3\alpha^2+\\
    &&\frac{(r^2-a^2)c^2}{\alpha^2(r^2+a^2)^3}. \label{21}
\end{eqnarray}
Now we can analyze all these components of the Riemann and Ricci tensors at the origin ($r\rightarrow 0$):
\begin{eqnarray}
    &&R^{01}{}{}_{01} = -\frac{\alpha^2[r_{HNS}^3+2a^3]}{2a^3}+\frac{c^2}{\alpha^2a^4}; \label{26}\\
    &&R^{02}{}{}_{02}=R^{03}{}{}_{03}=R^{23}{}{}_{23}=0; \\
    &&R^{12}{}{}_{12}=R^{13}{}{}_{13}= \frac{\alpha^2[r_{HNS}^3-a^3]}{a^3}-\frac{c^2}{\alpha^2a^4}; \label{22}\\
    &&R^{0}{}_{0}=-\frac{\alpha^2(2a^5+a^2r_{HNS}^3)}{2a^5}+\frac{c^2}{\alpha^2a^4}; \label{23}\\
    &&R^{1}{}_{1}=\frac{\alpha^2(3a^2r_{HNS}^3-6a^5)}{2a^5}-\frac{c^2}{\alpha^2a^4}; \label{24}\\
    &&R^{2}{}_{2} = R^{3}{}_{3} = \alpha^2\left(\frac{r_{HNS}^3}{a^3}+2\right)-3\alpha^2-\frac{c^2}{\alpha^2a^4}. \label{25}
\end{eqnarray}

Therefore, as we have seen in section (\ref{CI}), all these components are finite if $a\neq 0$. Now that we have verified the regularity of the solution, let's analyze the energy conditions associated with the stress-energy tensor through the solution of the Einstein equations.

\subsection{\textit{Stress-Energy Tensor and Energy Conditions of the Regular Charged Black String}}

To solve the Einstein equations, we need to determine the components of the Einstein tensor and the stress-energy tensor. We can find the components of the Einstein tensor from the Ricci tensor and the Ricci scalar:
\begin{eqnarray}
    \nonumber G^{0}{}_{0} &=& \frac{\alpha^2[3(r^2+a^2)^{5/2}-a^2((r^2+a^2))^{3/2}+2r_{HNS}^3)]}{(r^2+a^2)^{5/2}}+\\
    &&\frac{(2a^2-r^2)c^2}{\alpha^2(r^2+a^2)^3}; \label{26}\\
    G^{1}{}_{1} &=& \frac{3\alpha^2r^2}{(r^2+a^2)}-\frac{c^2r^2}{\alpha^2(r^2+a^2)^3}; \label{27}
\end{eqnarray}
\begin{eqnarray}
    \nonumber G^{2}{}_{2} = G^{3}{}_{3} &=& 3\alpha^2-\frac{a^2\alpha^2}{(r^2+a^2)}\left(\frac{r_{HNS}^3}{2 (r^2+a^2)^{3/2}}+1\right)+\\
    &&\frac{c^2r^2}{\alpha^2(r^2+a^2)^3}. \label{28}
\end{eqnarray}

On the other hand, the components of the stress-energy tensor are given by:
\begin{equation}\label{29}
    T^0{}_0=-\rho;\, T^{1}{}_{1}=p_{\|};\, T^{2}{}_{2}=T^{3}{}_{3}=p_{\bot}.
\end{equation}
This result is only valid in regions where $g(r)\geq 0$, that is, outside the outer horizon or inside the inner horizon, as we will also have two horizons here. Between the two horizons, there must be an inversion of the timelike and spacelike characteristics, as $g(r)<0$, which implies the change $T^0{}_0=p_{\|}$ and $T^{1}{}_{1}=-\rho$.

Using the Einstein equations with a cosmological constant, we can determine the components of the stress-energy tensor outside the outer horizon or inside the inner horizon.
\begin{eqnarray}
    &&\rho = \frac{\alpha^2a^2((r^2+a^2)^{3/2}+2r_{HNS}^3)}{8\pi (r^2+a^2)^{5/2}}+\frac{(r^2-2a^2)c^2}{8\pi \alpha^2(r^2+a^2)^3}; \label{30}\\
    &&p_{\|} = -\frac{3\alpha^2a^2}{8\pi (r^2+a^2)}-\frac{r^2c^2}{8\pi \alpha^2(r^2+a^2)^3}; \label{31}\\
    &&p_{\bot} = - \frac{a^2\alpha^2(2(r^2+a^2)^{3/2}+r_{HNS}^3)}{16\pi (r^2+a^2)^{5/2}}+\frac{r^2c^2}{8\pi \alpha^2(r^2+a^2)^3}. \label{32}
\end{eqnarray}
Now let's examine the energy conditions, particularly the null energy condition (NEC). According to the NEC, in order for it to hold, we must have ($\rho+p_{\|}$). Let's calculate $\rho+p_{\|}$:
\begin{equation}\label{33}
\rho + p_{\|} = -\frac{2a^2\alpha^2}{8\pi (r^2+a^2)^2}g(r),
\end{equation}

where $g(r)$ is the same as defined in equation (\ref{11}). We observe that the right-hand side of equation (\ref{33}) consists of quadratic terms, which are always positive, in addition to the function $g$ itself. Since we are considering the region outside the external horizon or inside the internal horizon, in these regions, we should have $g(r)>0$. Consequently, as the entire expression is multiplied by $-1$, we must have $\rho+p_{\|}<0$ for all $r$ in these regions. Therefore, the null condition is violated, similar to the regularization of the Reissner-Nordstrom solution \cite{Franzin:2021vnj}.

Considering the region between the horizons, we only need to change the sign of the result in equation (\ref{33}) for the null condition. However, in this case, since $g(r)<0$, the NEC will continue to be violated. When we consider the exact position of either of the horizons, we have $g(r)=0$, meaning that $\rho+p_{\|}=0$.

Now, we can perform a more specific analysis of the stress-energy tensor since we have already determined the functions that specify its components, namely the energy density $\rho(r)$, the radial pressure $p_{\|}(r)$, and the lateral pressure $p_{\bot}$. We find that the non-zero components of $T^{\mu}{}_{\nu}$ are a combination of a term that does not depend on the charge density (an $c$-independent term), which corresponds exactly to the solution of the neutral regular black string as described in \cite{Lima:2022pvc}, and an additional term that depends on the charge. Our objective now is to find a source that precisely reproduces these components of the stress-energy tensor as determined from equations (\ref{29}), (\ref{30}), (\ref{31}), and (\ref{32}).

\subsection{\textit{Sources for the regularized metric}}
The first step is to define which Lagrangians should be included in the Einstein-Hilbert action of our solution. As seen in \ref{Intro}, the most common combination of sources for black bounce solutions is that of a scalar field with nonlinear electrodynamics.  Therefore, it is reasonable to assume the following action for the charged black string:
\begin{equation}\label{34}
S=\int d^4x \sqrt{-g}[R-2\Lambda - 16\pi (\epsilon g^{\mu\nu}\partial_{\mu}\phi\partial_{\nu}\phi+V(\phi))-16\pi L(F)],
\end{equation}

In this expression, $\phi(r)$ refers to the scalar field, which can represent either the usual scalar field or the phantom field depending on the parameter $\epsilon$. Specifically, when $\epsilon=1$, it corresponds to the usual scalar field, whereas $\epsilon=-1$ corresponds to the phantom field. The function $V(\phi)$ represents the potential associated with the scalar field, and $L(F)$ is the Lagrangian related to nonlinear electrodynamics, which depends on $F=F^{\mu\nu}F_{\mu\nu}/4$ (it is worth noting that the usual Maxwell Lagrangian is recovered when $L=F$). The equations of motion associated with the action (\ref{34}) are as follows:
\begin{eqnarray}
    &&R^{\mu}{}_{\nu}-\frac{1}{2}\delta^{\mu}_{\nu}R+\delta^{\mu}_{\nu}\Lambda=8\pi([T_{\phi}]^{\mu}{}_{\nu} +[T_{NED}]^{\mu}{}_{\nu}), \label{35} \\
    &&2\epsilon \nabla_{\mu}\nabla^{\mu}\phi=\frac{dV}{d\phi}, \\
    &&\nabla_{\mu}[L_F F^{\mu\nu}]=\frac{1}{\sqrt{-g}}\partial_{\mu}[\sqrt{-g}L_F F^{\mu\nu}]=0,
\end{eqnarray}
where $L_F=\partial L/\partial F$, and the components of the scalar field and NED stress-energy tensor are, respectively:
\begin{eqnarray}
    &&[T_{\phi}]^{\mu}{}_{\nu}=2\epsilon\partial^{\mu}\phi\partial_{\nu}\phi-\delta^{\mu}_{\nu}(\epsilon\partial^{\alpha}\phi\partial_{\alpha}\phi+V(\phi)), \\
    &&{}[T_{NED}]^{\mu}{}_{\nu}=L_F F^{\mu\alpha}F_{\nu\alpha}-\delta^{\mu}_{\nu}L(F).
\end{eqnarray}

For simplicity, we will work with the metric using the function $g(r)$ and replace it with the expression defined in (\ref{11}) at the end. We also have $F_{\mu\nu}=\partial_{\mu}A_{\nu}-\partial_{\nu}A_{\mu}$, where we will use the Ansatz for $A_{\mu}$ as found in \cite{Hendi:2013mka}: $A_{\mu}=h(r)\delta^{0}_{\mu}$, since our problem does not involve rotation. The equations of motion can then be written as:

\begin{eqnarray}
    \nonumber &&\frac{(r^2+2a^2)g(r)}{(r^2+a^2)^2}+\frac{rg'(r)}{(r^2+a^2)}+\Lambda=-8\pi[\epsilon g(r)\phi'(r)^2+ \\
    &&V(r)+L_Fh'(r)^2+L(r)], \label{50}\\
    \nonumber &&\frac{r^2g(r)}{(r^2+a^2)^2}+\frac{rg'(r)}{(r^2+a^2)}+\Lambda=-8\pi[-\epsilon g(r)\phi'(r)^2+ \\
    &&V(r)+L_Fh'(r)^2+L(r)], \label{51}\\
    \nonumber &&\frac{a^2g(r)}{(r^2+a^2)^2}+\frac{rg'(r)}{(r^2+a^2)}+\frac{g''(r)}{2}+\Lambda=-8\pi[\epsilon g(r)\phi'(r)^2+ \\
    &&V(r)+L(r)], \label{52}\\
    &&2\epsilon(g(r)\phi''(r)+g'(r)\phi'(r))+\frac{4\epsilon rg(r)\phi'(r)}{(r^2+a^2)}=\frac{V'(r)}{\phi'(r)}, \label{53}\\
    &&[(r^2+a^2)L_Fh'(r)]'=0. \label{54}
\end{eqnarray}

The last equation is the simplest, as the term in brackets must be constant for its derivative to be zero. Thus, we have:
\begin{equation}\label{55}
    h'(r)=\frac{C}{L_F(r^2+a^2)},
\end{equation}
where $C$ is a constant. 




To calculate $h(r)$, we need to first determine $L_F(r)$. This can be done by isolating $L_F$ in equation (\ref{50}), substituting $h'(r)$ with (\ref{55}), and replacing $L(r)$ according to equation (\ref{52}):
\begin{equation}\label{59}
L_F(r)=\frac{16\pi \alpha^2 C^2(r^2+a^2)}{4c^2(r^2-a^2)+3\alpha ba^2\sqrt{r^2+a^2}}.
\end{equation}
Substituting (\ref{59}) into (\ref{55}), we can integrate it to determine $h(r)$:
\begin{equation}
h(r)=\frac{1}{16\pi C\alpha^2}\left[\frac{3\alpha b r}{\sqrt{r^2+a^2}}-\frac{4c^2r}{r^2+a^2}\right]+D.
\end{equation}
where $D$ is a constant. We note that  $C$ and $D$ must depend on the other constants. Let us fix them by considering the correct limits. First we note that, when $a\to 0$ we must get the usual charged black string solution $h=-c/(\alpha r)$ \cite{Lemos:1995cm}. Therefore $4\pi\alpha C(a=0)=c$ and $D(a=0)=3b/(4c)$. In the limit $c\to0$, we must obtain the source for the neutral black string and $C(c=0)$ must a non zero constant. We will choose $D(c=0)=0$. A good choice therefore is given by $4\pi\alpha C=(c^2+\alpha^2 a^2)/\sqrt{c^2+\alpha^4 a^4}$ and $D=3bc/(4\sqrt{c^4+\alpha^2a^2})$. 

To calculate $L(r)$, we need to determine $V(r)$ first. Before that, we need to calculate $\phi(r)$, which can be done by comparing the expression $V(r)+L_Fh'(r)^2+L(r)$ in equations (\ref{50}) and (\ref{51}).
\begin{equation}
    \phi'(r)=\frac{ia}{\sqrt{8\pi\epsilon}(r^2+a^2)},
\end{equation}
where $i=\sqrt{-1}$. In order for $\phi(r)$ to be real, it is necessary to set $\epsilon=-1$ to eliminate $i$. This implies that we need to have a ghost field in our solution. Therefore:
\begin{equation}\label{60}
    \phi'(r)=\frac{1}{\sqrt{8\pi}}\frac{a}{(r^2+a^2)} \rightarrow \phi(r)=\frac{1}{\sqrt{8\pi}}\text{tg}^{-1}\left(\frac{r}{a}\right)+\phi_0,
\end{equation}
where $\phi_0$ is an integration constant. This result is consistent with our analysis that wormhole solutions are well described by ghost fields. Furthermore, the expression (\ref{60}) is identical to that found in \cite{Bronnikov:2021uta} for the Simpson-Visser regularization of the Reisnerr-Nordstrom solution. Now, it is sufficient to substitute the result of (\ref{60}) into (\ref{53}) to determine $V(r)$:
\begin{equation}\label{64}
V(r)=\frac{2a^2}{8\pi}\left[\frac{b}{5\alpha(r^2+a^2)^{5/2}}+\frac{\alpha^2}{r^2+a^2}-\frac{c^2}{3\alpha^2(r^2+a^2)^3}\right].
\end{equation}
We can express $r$ as a function of $\phi$ to find $V(\phi)$:
\begin{eqnarray}
 \nonumber 8\pi V(\phi)&=&2\alpha^2\cos^2(\sqrt{8\pi}\phi)+\frac{2b}{5\alpha a^3}\cos^5(\sqrt{8\pi}\phi)-\\
&&\frac{2c^2}{3\alpha^2a^4}\cos^6(\sqrt{8\pi}\phi).
\end{eqnarray}
Once the potential $V(r)$ is determined, we can obtain $L(r)$ using equation (\ref{52}):
\begin{equation}\label{67}
    L(r)=\frac{1}{\alpha^2 8\pi(r^2+a^2)^3}\left[c^2\left(\frac{5a^2}{3}-r^2\right)-\frac{9\alpha ba^2\sqrt{r^2+a^2}}{10}\right].
\end{equation}
We could find $L(F)$ since $F=-h'(r)^2/2$ and $h'(r)$ is determined in (\ref{55}). However, this solution is not straightforward because it would require isolating $r$ as a function of $F$ to determine $r(F)$ and then substituting it into (\ref{67}) to determine $L(F)$. However, this is not feasible as the solution for $r(F)$ is not analytical.

Thus, we have obtained all the necessary functions for the analysis of the sources in our solution. 

\section{Circular Orbits of Regularized Charged Black String}

One of the commonly performed analyses for black holes is regarding their possible circular orbits, both for massive particles and photons. For this analysis, we will use a method analogous to that of \cite{Franzin:2021vnj}, where we evaluate the effective potential energy of the system ($V_{\text{eff}}(r)$). To find the possible circular orbits for photons around the $z$-axis, we must use the condition that these orbits should minimize or maximize the effective potential energy. In the case of a minimum point, we have a stable equilibrium point ($V_{\text{eff}}''>0$), while in the case of a maximum point, it will be unstable ($V_{\text{eff}}''<0$) \cite{Jefremov:2015gza}.

The effective potential energy can be defined based on the metric (\ref{10}):
\begin{eqnarray}
    \nonumber \left(\frac{ds}{d\tau}\right)^2&=&-g(r)\left(\frac{dt}{d\tau}\right)^2 + g(r)^{-1}\left(\frac{dr}{d\tau}\right)^2 + (r^2+a^2)\left(\frac{d\varphi}{d\tau}\right)^2 +\\
    &&\alpha^2(r^2+a^2)\left(\frac{dz}{d\tau}\right)^2, \label{38}
\end{eqnarray}
where $\tau$ is the proper time.

For simplicity, let's consider only the equatorial region $z=0$. Additionally, we define the term on the left-hand side of equation (\ref{38}) as $\epsilon$. Thus, for null orbits (associated with massless particles), we set $\epsilon=0$, while for massive particles, we have $\epsilon=-1$. We can also define the energy per unit mass and the angular momentum per unit mass as follows:
\begin{equation}
    E=g(r)\frac{dt}{d\tau};\, L=(r^2+a^2)\frac{d\varphi}{d\tau}.
\end{equation}
Thus, we have:
\begin{equation}
\left(\frac{dr}{d\tau}\right)^2 = E^2 + \left(\epsilon-\frac{L^2}{r^2+a^2}\right)g(r).
\end{equation}

Following the same definition used in \cite{Simpson:2018tsi, Franzin:2021vnj}, we have:
\begin{equation}
\left(\frac{dr}{d\tau}\right)^2 \equiv E^2 - V_{\text{eff}}(r),
\end{equation}
therefore:
\begin{equation}\label{72}
V_{\text{eff}}(r)=\left(\frac{L^2}{r^2+a^2}-\epsilon\right)g(r).
\end{equation}
Let's evaluate each case, first for massless particles and then for massive particles. 

\subsection{\textit{Circular orbits for photons}}
In this case, we set $\epsilon=0$ in equation (\ref{72}). To find the position $r$ of the circular orbit, let's determine the critical points of this energy effective, that is, we want to find the values of $r$ such that $V'_{\text{eff}}(r)=0$. Hence, we have:
\begin{equation}
V'_{\text{eff}}(r)=\frac{rL^2}{(r^2+a^2)^{5/2}}\left(\frac{3b}{\alpha}-\frac{4c^2}{\alpha^2\sqrt{r^2+a^2}}\right).
\end{equation}

Discarding the solution $r=0$ since we cannot have an orbit with zero radius, we have:
\begin{equation}\label{74}
\sqrt{r^2+a^2} = \frac{4c^2}{3\alpha b} \rightarrow r = \pm \sqrt{\frac{16c^4}{9\alpha^2b^2}-a^2},
\end{equation}

where the signs of $r$ are related to the "universe" we are dealing with in the solution. For simplicity, we can choose the upper universe, which has a positive sign. In this case, the solution is:
\begin{equation}\label{54}
   r\equiv r_{P} = r_{PCS}\sqrt{1-\frac{a^2}{r_{PCS}^2}},
\end{equation}
where $r_{PCS}=4c^2/3\alpha b$ is the value of the circular orbit of the charged solution for massless particles in the non-regular case ($a=0$) \cite{Lemos:1995cm}. We see that, unlike the uncharged case, it is possible to have circular orbits as long as $a<r_{PSC}$. We can verify whether this orbit is stable or unstable by analyzing $V''_{\text{eff}}(r)$:
\begin{eqnarray}
    \nonumber V''_{\text{eff}}(r)&=&L^2\left(\frac{(r^2+a^2)^{3/2}-5r}{(r^2+a^2)^{7/2}}\right)\left(\frac{3b}{\alpha}-\frac{4c^2}{\alpha^2\sqrt{r^2+a^2}}\right)+\\
    &&L^2\frac{4c^2r^2}{\alpha^2(r^2+a^2)^4}.
\end{eqnarray}

Now we evaluate the value of $V''_{\text{eff}}(r)$ at the position of the circular orbit given in (\ref{54}):
\begin{equation}\label{76}
   V''_{\text{eff}}(r_{P})= \frac{4L^2c^2}{\alpha^2r_{PCS}^6}\left(1-\frac{a^2}{r_{PCS}^2}\right),
\end{equation}
where it is clear that under the condition for this orbit to exist ($a<r_{PCS}$), the sign of $V''_{\text{eff}}$ will always be positive, which implies that this orbit is stable. This result is consistent with the non-regular solution ($a=0$), where this orbit should be stable \cite{Lemos:1995cm}.

We can see that this result is consistent with what was found in \cite{Lima:2022pvc}, which deals with the regular uncharged black string. If we consider the case where $c=0$ in equation (\ref{74}), we would have $r^2=-a^2$, which would result in a complex value for $r$. Therefore, a circular orbit for photons is only possible if there is a charge ($c\neq 0$).
\subsection{\textit{Circular orbits for massive particles}}
For the case of massive particles, we must have $\epsilon=-1$ in equation (\ref{72}). To simplify the calculations, let's use the variable $\overline{r}$ again, as defined in section \ref{EHSG}. In this case, the derivative of $V_{\text{eff}}(\overline{r})$ is given by:
\begin{eqnarray}
    V'_{\text{eff}}(\overline{r})=2\left[\alpha^2-\frac{b}{\alpha\overline{r}^3}+\frac{c^2}{\alpha^2\overline{r}^4}\right]+\frac{(L^2+\overline{r}^2)}{\overline{r}^5}\left[\frac{3b}{\alpha}-\frac{4c^2}{\alpha^2\overline{r}}\right].
\end{eqnarray}
Now, let's impose the condition for a circular orbit ($V'_{\text{eff}}(\overline{r})=0$) and then isolate $L$:
\begin{equation}\label{78}
    L(\overline{r})=\sqrt{\frac{2\overline{r}^2c^2-2\alpha^4\overline{r}^6-b\alpha\overline{r}^3}{3b\alpha\overline{r}-4c^2}};
\end{equation}
Once again, we see consistency with the results obtained in \cite{Lima:2022pvc} for the case where $c=0$, as in this case, it is not possible to have circular orbits since the angular momentum would have to be complex, which is evident in equation (\ref{78}). In our case, to determine the orbit known as the ISCO (inner stable circular orbit), we need to find the minimum value of $L(\overline{r})$, that is, find $\overline{r}$ such that $\partial L/\partial\overline{r}=0$. This occurs for the solution of the following equation:
\begin{equation}
    15\alpha^5b\overline{r}^5-24\alpha^4c^2\overline{r}^4+3\alpha^2b^2\overline{r}^2-9\alpha bc^2\overline{r}+8c^4=0.
\end{equation}
This equation does not have an analytical solution since it is a fifth-degree equation. Therefore, we need the numerical values of each constant to determine the numerical solutions for $\overline{r}$.
\section{Conclusion}

This paper investigates the solution proposed by Lemos in \cite{Lemos:1995cm} for a charged black string. The regularization method of Simpson-Visser is applied, involving the modification $r\rightarrow \sqrt{r^2+a^2}$ in the non-regular solution. The obtained solution represents both a regular black string and a traversable wormhole, depending on the parameter $a$. It can have up to four horizons, corresponding to different universes and regions of $r$. The horizons, surface gravity, and Hawking temperature are determined based on the original Lemos solution.

The regularity of the solution is assessed by examining curvature invariants and tensors, such as the Kretschmann scalar and the Ricci scalar. No singularities are found at the origin as long as $a\neq 0$, consistent with other black bounce solutions. The analysis also confirms the consistency of curvature tensors with the uncharged regular black string. Energy-momentum tensor and energy conditions are analyzed, with the null energy condition being violated throughout the space.

Two types of sources are considered: a scalar field and nonlinear electrodynamics (NED). The scalar field results are consistent with previous works, requiring a ghost-type field for reality. However, the treatment of the electrodynamics differs from typical approaches in black bounce sources, involving an electric source instead of magnetic charge.

Stable and unstable circular orbits for massless and massive particles are investigated. Stable orbits exist for massless particles (photons) within certain parameter constraints. For massive particles, finding the smallest stable circular orbit (ISCO) requires solving a fifth-degree equation numerically. The absence of an ISCO is observed in the uncharged case, consistent with previous findings.

Future work includes exploring the magnetic case, analyzing causal structures using Penrose diagrams, and studying rotating solutions.

\section*{Acknowledgements}

The authors would like to thank Conselho Nacional de Desenvolvimento Científico e Tecnológico (CNPq), Fundação Cearense de Apoio ao Desenvolvimento Científico e Tecnológico
(FUNCAP) and Coordenação de Aperfeiçoamento de Pessoal de Nível Superior - Brasil (CAPES) for finantial support. 

\subsubsection*{Note added.}

At the time this paper appeared as a preprint, we became aware of Bronnikov et al. work on arXiv studying the same subject we address here \cite{Bronnikov:2023aya}.

\end{document}